\documentclass[useAMS,usenatbib]{mn2e}
\usepackage{graphicx,amsmath,multirow,amssymb}
\usepackage{natbib}
\newcommand{\comment}[1]{}

\def\simgt{\lower.5ex\hbox{$\; \buildrel > \over \sim \;$}}
\def\simlt{\lower.5ex\hbox{$\; \buildrel < \over \sim \;$}}

\title[Dust in the winds of low--metallicity AGBs]{Dust formation in the winds of AGBs: 
the contribution at low metallicities}
\author[Di Criscienzo et al.]{M. Di Criscienzo$^{1,2}$, F. Dell'Agli$^{3}$, P. Ventura$^1$, R. Schneider$^1$, 
R. Valiante$^1$,  
\newauthor
F. La Franca$^{4}$, C. Rossi$^{3}$, S. Gallerani$^5$, R. Maiolino$^6$  \\
$^1$INAF -- Osservatorio Astronomico di Roma, Via Frascati 33, 00040, Monte Porzio Catone (RM), Italy \\
$^2$INAF -- Osservatorio Astronomico di Capodimonte, Salita Moiarello 16, 80131, NApoli, Italy \\
$^{3}$Dipartimento di Fisica, Universit\`a di Roma ``La Sapienza'', P.le Aldo Moro 5, 00143, 
Roma, Italy \\
$^{4}$Dipartimento di Fisica, Universit\`a di Roma ``Roma Tre'', P.le Aldo Moro 5, 00143,  Roma, Italy \\
$^{5}$Scuola Normale Superiore, Piazza dei Cavalieri 7, 56126 Pisa, Italy\\
$^{6}$Cavendish Laboratory, University of Cambridge, 19 J.J. Thomson Ave., Cambridge CB3 OHE, UK}

\begin{document}

\date{Accepted, Received; in original form }

\pagerange{\pageref{firstpage}--\pageref{lastpage}} \pubyear{2012}

\maketitle

\label{firstpage}

\begin{abstract}
We present new models for the evolution of stars with mass in the range 
$1 M_{\odot} \leq M \leq 7.5 M_{\odot}$, followed from the pre--main--sequence through
the asymptotic giant branch phase, until most of their envelope is lost via stellar winds.
The metallicity adopted is $Z=3\times 10^{-4}$ (which, with an $\alpha-$enhancement of 
$+0.4$, corresponds to $[Fe/H]=-2$). Dust
formation is described by following the growth of dust grains of various types as
the wind expands from the stellar surface. 

Models with mass $M\geq 3 M_{\odot}$ experience Hot Bottom Burning, thus maintaining the
surface C/O below unity. Unlike higher $Z$ models, the scarcity of silicon available in the 
envelope prevents the formation of silicates in meaningful quantities, 
sufficient to trigger the acceleration of the wind via radiation pressure on the dust 
grains formed. No silicate formation occurs below a threshold metallicity of 
$Z=10^{-3}$.

Low--mass stars, with $M\leq 2.5 M_{\odot}$ become carbon stars, forming solid carbon
dust in their surroundings. The total dust mass formed depends on the uncertain extent of 
the inwards penetration of the convective envelope during the Third Dredge--Up episodes 
following the Thermal Pulses. However, provided that a minimum abundance of carbon of 
$X(C) \sim 5\times 10^{-3}$ is reached in the envelope, the results turn out to be
fairly independent of the parameters used. Carbon grains have sizes 
$0.08$ $\ mu m < a_C < 0.12$ $\mu$m and the total amount of dust formed (increasing with the
mass of the star) is $M_C=(2-6)\times 10^{-4}$ $ M_{\odot}$.

Our results imply that AGB stars with $Z=3\times 10^{-4}$ can only contribute to carbon
dust enrichment of the interstellar medium on relatively long timescales, $ > 300$~Myr, 
comparable to the evolutionary time of a $3 M_{\odot}$ star. At lower metallicities the 
scarcity of silicon available and the presence of Hot Bottom Burning
even in M$< 2$ M$_{\odot}$, prevents the formation of silicate and 
carbon grains. We extrapolate our conclusion to more metal--poor environments, and
deduce that at $Z < 10^{-4}$ dust enrichment is mostly due to metal condensation in 
supernova ejecta.   
\end{abstract}


\section{Introduction}
The importance of intermediate mass stars on the asymptotic giant branch (AGB) and 
super asymptotic giant branch (super--AGB) as source of cosmic dust is currently a highly debated topic. 
Recent chemical evolutionary models \citep{valiante09, valiante11} have shown that AGB stars give 
important contributions to dust enrichment even at early cosmic epochs, when traditionally 
core-collapse supernovae (CCSNe) have been proposed as the main source of dust 
\citep{todini01, nozawa03, bianchi07, dwek07}.
The recent detection of about $(0.4 - 0.7) M_{\odot}$ of cold dust in the supernova 1987A 
by the Herschel satellite \citep{matsuura11} confirms the results expected by theoretical
models of dust condensation in supernova ejecta. Yet, only a relatively small fraction of the
newly formed dust, probably less than 20\%, is able to survive the passage of the reverse shock 
and be injected in the surrounding interstellar medium (ISM), effectively contributing to dust
enrichment of the host galaxy \citep{bianchi07, nozawa07}.
Indeed, Spitzer observations of young supernova remnants have provided unabiguous evidence of
dust formation and survival in the ejecta, with typical masses in the range 
$(1-5)\times 10^{-2} M_{\odot}$ \citep{rho09}. These figures can be considered as an indication
of the effective dust yields of massive stars.

At lower stellar masses ($< 8 M_{\odot}$) grain condensation occurs during the AGB and super-AGB
phases and the resulting dust mass depends on the progenitor stellar mass and metallicity through
complex processes which affect the chemical composition and physical conditions in the
stellar atmospheres. 

Previous investigations, based on synthetic stellar models, have provided
grids of AGB dust yields predicting dust masses in the range $(10^{-4} - 10^{-2}) M_{\odot}$
(Ferrarotti \& Gail 2006). With the aim to make further progresses following the proper stellar
evolutionary sequences, we have recently computed the dust masses released by stars with
masses in the range $1 M_{\odot} \le M \le 8 M_{\odot}$ and metallicities of $Z = 0.001 
\, (5\times 10^{-2} Z_{\odot})$ and $Z = 0.008 \, (0.4 Z_{\odot})$. We find that the
both the total mass of dust released and its chemical composition are a strong function
of the initial stellar mass and metallicity (Ventura et al. 2012a, 2012b) and that, in
the most optimistic cases, the typical dust masses released by AGB stars can reach up
to $7 \times 10^{-3} M_{\odot}$. 
While the relative importance of AGB stars and SN as sources of dust will ultimately
depend on the star formation history and stellar initial mass function 
\citep{valiante09, valiante11}, in order to understand the contribution of AGB stars in 
early cosmic dust enrichment we need to extend our previous models to lower initial 
metallicities. 

In this paper, we calculate a new sequence of models with metallicity 
$Z=0.0003 \, (1.5\times 10^{-2} Z_{\odot})$, evolved through the whole AGB phase. Our
results imply that at these low initial metallicities, the production of dust is strongly
suppressed, suggesting that in cosmic environments with metallicities $Z < 5\times 10^{-3} Z_{\odot}$
dust enrichment is entirely dominated by supernovae.

\section{The model}
\label{refmodel}
The thermal pulses (TP) phase is experienced by stars of intermediate mass 
($M < (8-9) M_{\odot}$), after helium is consumed in the core (see e.g. \citet{herwig05}
for an exhaustive description of these phases). Helium and hydrogen burning 
occur in two shells above the degenerate core. This evolutionary phase is
commonly known as Asymptotic Giant Branch (AGB).
In the most massive models, whose core mass exceeds $\sim 1.1M_{\odot}$, carbon ignition
occurs in an off--center, degenerate region. This carbon pulse favors the development
of a convective flame, that moves inwards to reach the stellar center, leading to the
formation of a core composed of oxygen and neon \citep{garciaberro97, ritossa96, ritossa99}. 
These stars also experience a series of thermal pulses, during the so called Super 
Asymptotic Giant Branch evolution \citep{siess07, siess10}.

The surface chemistry of these stars may be altered by two mechanisms: Hot Bottom Burning 
(HBB) and Third Dredge--Up (TDU). HBB is active in massive AGBs, and consists in the 
ignition of proton--capture nucleosynthesis at the bottom of the convective envelope; 
it is activated when the temperature of this region reaches $\sim 5\times 10^7$ K. TDU 
consists in the inwards penetration of the envelope after each TP; the bottom of the 
surface convection zone may reach regions processed by $3\alpha$ nucleosynthesis, enriched 
in carbon and oxygen.

The modification of the surface chemistry produced by HBB and TDU are different. TDU is
accompanied by the increase in the surface abundance of carbon, and, to a smaller extent, 
of oxygen. The effects of HBB reflect proton--capture nucleosynthesis, and thus depend on the 
temperature at which HBB occurs. For temperatures below $\sim 8 \times 10^7$K the main effect is 
the production of nitrogen at the expenses of carbon, whereas at higher temperatures oxygen 
destruction also occurs.

Due to the large mass loss rates experienced, and the cool temperatures in their expanded,
low--density envelopes, the winds of AGBs are a favorable site for dust production, via
condensation of gas molecules into solid grains. The investigations by \citet{fg01, fg02, fg06} 
set the theoretical framework to model dust formation in expanding AGB winds, and showed how 
the formation of carbonaceous species (essentially solid carbon and SiC) and silicates 
(olivine, pyroxene, quartz) depends on the surface chemistry of AGBs, particularly on the 
C/O ratio. These works are based on a synthetic description of the AGB phase, where 
core mass, temperature at the bottom of the convective zone, extension of TDU, are all
imposed a priori, based on the observational calibration. This approach
has the clear advantage of allowing one to follow the whole AGB evolution with small
computational efforts; the intrinsic limit of this description is that HBB cannot be
described self--consistently. 

In our previous investigations 
(Ventura et al. 2012a,b, hereafter paper I and paper II, respectively) we presented 
new models where the AGB evolution was calculated
by means of the integration of the whole stellar structure. We also extended the calculations
to the super--AGB regime, ignored in previous works. Our analysis showed that the kind of dust 
formed around AGBs depends on the initial mass of the star: models with $M\geq 3M_{\odot}$,
whose surface chemistry is dominated by HBB, produce silicates, whereas their lower mass
counterparts produce carbon--type dust. These investigations also showed an interesting
trend with metallicity: in higher--$Z$ models the production of silicates is favored, not
only by the larger amount of silicon available in the envelope, but also because the HBB 
experienced is weaker, thus inhibiting the possibility of surface oxygen destruction.  

In what follows we further extend the model to lower initial metallicities and discuss the
implications for the mass and composition of dust yields by intermediate mass stars.

Dust grains are assumed to form via condensation of gas molecules in the wind of AGBs.
The description of this process requires knowledge of the evolution of the main physical
and chemical properties of the central object, and a model for the thermodynamical and
chemical structure of the wind. We provide here a brief description of the stellar and
wind modeling, addressing the interested reader to \citet{paperI} for a thorough
presentation of the model.

\subsection{Stellar evolution modelling}
The stellar models presented in this work have an initial metallicity $Z=3\times 10^{-4}$.
The mixture is assumed to be $\alpha-$enhanced, with $[\alpha/Fe]=+0.4$. This 
corresponds to an iron content $[Fe/H]=-2$. The evolutionary sequences were calculated
by means of the ATON stellar evolution code, in the version described in \citet{ventura98}.

The range of masses involved is limited to $M\leq 7.5M_{\odot}$, because more massive objects 
undergo core--collapse SN type II explosion, thus not experiencing any AGB 
phase\footnote{This limit is indeed
partly dependent on the assumption of some extra mixing from the border of the convective
core during the Main Sequence phase. If overshooting is neglected, the highest mass not
undergoing SNII explosion is $\sim 9.5M_{\odot}$}. The evolutionary models were followed 
from the early pre main sequence phase, until the phase where almost all the external 
envelope is consumed. Models with masses below $2.5M_{\odot}$ develop a degenerate helium 
core while ascending the red giant branch, and undergo the helium flash. The evolution of 
these models was thus re--started from the horizontal branch, from a post--flash stage, 
before the beginning of the quiescent core helium burning.

The convective instability is modeled according to the Full Spectrum of Turbulence
(hereinafter FST) treatment by \citet{cm91}. In regions unstable to convection,
mixing of chemicals and nuclear burning are coupled via a diffusion--like equation,
following the scheme presented in \citet{cloutman}. The overshoot of convective eddies
within radiatively stable regions is described via an exponential decay of velocities 
from the formal border of convection; this is determined via the classic Schwarzschild 
criterium, stating that the zone of neutral stability, where buoyancy vanishes, corresponds
to the condition $\nabla_{\rm rad}=\nabla_{\rm ad}$ (where $\nabla_{\rm rad}$ and 
$\nabla_{\rm ad}$ are, respectively the logarithmic temperature gradients when
the energy is entirely transported by radiation, or for an adiabatic transformation). 
The e--folding distance $l$ of the
exponential decay of convective velocities is assumed to be $l=\zeta \times H_p$, where
$H_p$ is the pressure scale height, $H_p=P/(g\rho)$ and $\zeta$ is a measure of the extension 
of the overshoot region. Following the calibration given in \citet{ventura98}, we adopt
$\zeta=0.02$ in all the evolutionary phases preceding the thermal pulses phase. The possible
extension of the extra mixing during the AGB phase is still an open issue, largely debated 
in the literature \citep{karakas11, stancliffe11}. To limit the number of free parameters, here we only allow some extra 
mixing from the borders of the convective shell that forms during the thermal pulse, and 
discuss how the results obtained depend on the assumed value of $\zeta$.

Mass loss was modeled following the treatment by \citet{blocker95}. 

The majority of the nuclear reaction rates were taken from the NACRE compilation 
\citep{angulo}. The only exceptions are the $3\alpha$ cross sections \citep{fynbo},
the rate of the $^{12}C+\alpha$ reaction \citep{kunz}, and some proton--captures,
not relevant for the present investigation.

\subsection{Dust production and the stellar wind}
The scheme adopted to model the structure of the wind at a given phase during the AGB
evolution is the same as in paper I, based on the description given in the series
of papers by \citet{fg01, fg02, fg06}. Here we briefly recall the schematization adopted.

The wind is assumed to expand isotropically from the stellar surface. The physical quantities
characterizing the central star are luminosity ($L$), effective temperature ($T_{\rm eff}$),
radius ($R_*$), mass ($M$), mass loss rate ($\dot M$).

%%%%%Table rates%%%%%%
\begin{table*}
\begin{center}
\caption{Dust species considered in the present analysis, their formation reaction and
the corresponding key elements (see text).} 
\begin{tabular}{l|l|c|c}
\hline
\hline 
Grain Species & Formation Reaction & Key element & Sticking coefficient\\
\hline
Olivine & 2$x$Mg +2(1-$x$)Fe+SiO+3H$_2$O $\rightarrow$ Mg$_{2x}$Fe$_{2(1-x)}$SiO$_4$ + 3H$_2$ & Si & 0.1 \\ 
Pyroxene & $x$Mg +(1-$x$)Fe+SiO+2H$_2$O  $\rightarrow$ Mg$_{x}$Fe$_{(1-x)}$SiO$_3$ + 2H$_2$ & Si & 0.1 \\
Quartz & SiO + H$_2$O $\rightarrow$ SiO$_2(s)$ +H$_2$ & Si & 0.1 \\  
Silicon Carbide & 2Si + C$_2$H$_2$ $\rightarrow$ 2 SiC + H$_2$ & Si & 1 \\ 
Carbon & C $\rightarrow$ C$(s)$ & C & 1 \\
Iron & Fe $\rightarrow$ Fe$(s)$ & Fe & 1 \\
\hline 
\hline 
\end{tabular}
\end{center}
\label{tabrates}
\end{table*}
%%%%%Table rates%%%%%%
%Dust formation is allowed to occur when conditions are favorable to condensation of gas molecules into dust grains. 
The structure of the wind is described by the behavior of velocity,
temperature, density, optical depth, as a function of the radial distance from the surface 
of the star. 

The equation of momentum conservation determines the radial variation of velocity:

\begin{equation}
v{dv\over dr}=-{GM\over r^2}(1-\Gamma)
\label{eqvel}
\end{equation}

\noindent
where $\Gamma$ is the key--quantity that determines whether the wind is 
accelerated by radiation pressure. It is given by the ratio between the radiation pressure 
on the dust grains and the gravitational pull:

\begin{equation}
\Gamma={kL\over 4\pi cGM}
\label{eqgamma}
\end{equation}

\noindent
where $k$ expresses the opacity, as determined by the interaction of the various species of
dust grains and the electromagnetic radiation. In the equation describing the radial variation 
of velocity we neglect pressure forces, because these are negligible compared to the gravity term.
This assumption holds in the present treatment, because we do not consider any
shock structure of the outflow.

The description of the radial behavior of the wind is completed by the equation for the
radial decay of the optical depth:

\begin{equation}
{d \tau \over dr}=-\rho k {R_*^2\over r^2}
\label{eqtau}
\end{equation}

\noindent
The boundary condition for the optical depth is that it vanishes at infinity. We keep the 
velocity constant, and equal to the local sound speed, up to the region where dust begins 
to form. Whenever dust is formed, and the gas is accelerated by radiation pressure, the results are 
practically independent of the assumed initial value of velocity, because the asymptotic 
behavior turns out to depend only on the amount of dust formed. 

Finally, two additional equations describe the radial variation of density and temperature:

\begin{equation}
\dot M=4\pi r^2 \rho v
\label{eqmloss}
\end{equation}

\noindent
\begin{equation}
T^4={1\over 2}T_{\rm eff}^4 \left[ 1-\sqrt{1-{R_*^2\over r^2}}+{3\over 2}\tau \right]
\end{equation}

\noindent
The system of equations given so far is not closed, because the calculation of the 
opacity, $k$, requires knowledge of the kind of dust formed in the wind; it is therefore 
mandatory to follow the formation and growth of the dust grains of the various species.

Dust growth takes place by vapour deposition on the surface of some pre-formed seed nuclei, 
assumed to be nano--meter sized spheres. The precise chemical nature of these seed grains 
is not important in what follows. We consider various types of dust, depending on the 
surface chemical composition of the star. In oxygen--rich winds, we consider olivine, 
pyroxene, quartz and iron grains, whereas for C--rich environments we account for the 
presence of solid carbon, silicon carbide and iron grains. For each condensate, we define 
a key element, whose abundance is the minimum among all the elements necessary to form the 
corresponding dust aggregate. Silicon is the key species for silicon carbide, whereas iron 
and carbon are the key elements for iron dust and solid carbon. As for the silicates,
the key element is silicon, which is however replaced by oxygen when strong HBB conditions
favor a strong depletion of the surface oxygen content. All the species considered, with 
the corresponding condensation reactions and key elements, are listed in Table~1. 

The growth of the size of the dust grains of a given species $i$ is determined via a
competition between a production term ($J_i^{\rm gr}$), associated to the deposition of 
new molecules on already formed grains, and a destruction factor ($J_i^{\rm dec}$), 
proportional to the vapor pressure of the key species on the solid state:

\noindent
\begin{equation}
{d a_i \over dt}=V_{0,i}(J_i^{\rm gr}-J_i^{\rm dec})
\label{size}
\end{equation}

\noindent
where $V_{0,i}$ is the volume of the nominal molecule in the solid, and
the growth and destruction rates are given by

\begin{equation}
J_i^{\rm gr}=\alpha_i n_i v_{th,i}
\label{grrate}
\end{equation}
\noindent

\begin{equation}
J_i^{\rm dec}=\alpha_i v_{th,i} {P_{v,i}\over kT}
\label{decrate}
\end{equation}

\noindent
where $n_i$ is the number density of the key species in the wind, $v_{th,i}$ is the thermal
velocity of the same species, and $P_{v,i}$ is the vapor pressure of the key element 
over the solid state. As we see from Eq.~\ref{size}, \ref{grrate}, \ref{decrate}, the rate 
of growth of the dust grains is proportional to the sticking coefficients $\alpha_i$, 
which vary between zero and unity. The coefficients adopted for each species, taken from 
\citet{fg06}, are reported in Table 1.

\begin{figure}
\begin{minipage}{0.45\textwidth}
\resizebox{1.\hsize}{!}{\includegraphics{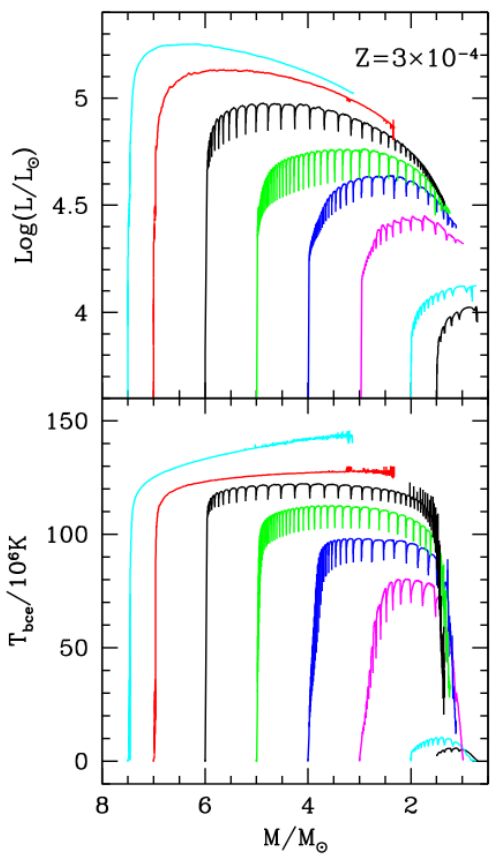}}
\end{minipage}
\vskip-10pt
\caption{Variation of the temperature at the bottom of the convective envelope ({\bf bottom})
and of the luminosity ({\bf top}) of models of initial mass in the range 
$1.5M_{\odot} \leq M \leq 7.5M_{\odot}$ evolved during the AGB phase. The total mass of
the star, reported in the abscissa, is decreasing during the evolution.
}
\label{fagb}
\end{figure}

\begin{table*}
\caption{Relevant properties of AGB models. The last column indicates the physical process
dominating the surface chemistry.}
\label{yields}
\begin{tabular}{cccccc}
\hline
\hline
$M/M_{\odot}$ & $\tau_{\rm evol}$ (Myr) & $M_{\rm core}/M_{\odot}$ & $T_{\rm bce}^{\rm max}$ ($10^6$K) & $\log(L/L_{\odot})$  \\
\hline
 & & & Z=$3\times 10^{-4}$ &  \\
\hline
1.0  & 5921   & .60  &  4    &  3.78  &  TDU \\
1.5  & 1065   & .68  &  4    &  3.98  &  TDU \\
2.0  &  805   & .72  & 11    &  4.13  &  TDU \\
2.5  &  455   & .78  & 38    &  4.27  &  TDU \\
3.0  &  294   & .84  & 79    &  4.42  &  HBB + TDU \\
3.5  &  207   & .87  & 89    &  4.54  &  HBB + TDU \\
4.0  &  155   & .90  & 98    &  4.63  &  HBB + TDU \\
4.5  &  120   & .93  & 100   &  4.70  &  HBB + TDU \\
5.0  &  96.0  & .97  & 111   &  4.75  &  HBB + TDU \\
5.5  &  79.0  & 1.00 & 117   &  4.85  &  HBB \\
6.0  &  67.0  & 1.05 & 121   &  4.97  &  HBB \\
6.5  &  57.0  & 1.16 & 124   &  5.05  &  HBB \\
7.0  &  49.5  & 1.25 & 127   &  5.12  &  HBB \\
7.5  &  43.5  & 1.34 & 141   &  5.20  &  HBB \\
\hline
\hline
\end{tabular}
\end{table*}

\section{The evolution of low--Z stellar models}
Models with initial metallicity $Z=3\times 10^{-4}$ and mass in the range 
$1M_{\odot} \leq M \leq 7.5M_{\odot}$ are evolved through the thermal pulses phase, until 
most of the mass in the envelope is lost. We find that, for this $Z$, $7.5M_{\odot}$ is 
the largest mass not undergoing supernova explosion. 

The physical properties of the models, with the maximum luminosity and temperature at the 
bottom of the envelope reached during the AGB evolution, are reported in Table \ref{yields}.

The top panel of Fig.~\ref{fagb} shows the variation of the luminosity of models 
of different mass, during the AGB phase. The evolution is plotted as a function of the mass
(decreasing with time), to better understand the main properties of the stars when most of 
the mass loss occurs.

More--massive models, with larger cores, evolve to larger luminosities. For each mass 
the overall energy release first increases during the early AGB, then decreases in the 
late phases. While the initial increase is due to the growing core mass, the following drop in
the luminosity is favoured by mass loss, which decreases the mass in the envelope.  
Because mass loss increases with luminosity \citep{blocker95}, the models 
evolving at large luminosity are also those losing mass at the highest rates.

\subsection{The evolution of high--mass AGB stars}
The bottom panel of Fig.~\ref{fagb} shows the evolution of the temperature at the bottom 
of the convective zone ($T_{\rm bce}$). A clear gap separates models with mass above $2M_{\odot}$ 
from their smaller mass counterparts. This is the signature of Hot Bottom Burning, 
operating when the core mass exceeds the threshold mass of $\sim 0.8M_{\odot}$, reached 
for initial masses $M>2.5M_{\odot}$. HBB is a consequence of the steepness of the 
temperature gradient: under these conditions the bottom of the
convective envelope partially overlaps with the CNO burning shell, thus allowing some
nuclear activity to take place within the outer convective zone. The pioneering investigations
by \citet{renzini81} and \citet{blocker91} showed that HBB ignition can be achieved
whenever a high--efficiency treatment of convection is adopted. \citet{vd05a} showed that 
HBB is easily obtained in models with $M\geq 3M_{\odot}$ when the FST treatment of convection
is used. HBB is a complex phenomenon, where the thermodynamic coupling between the internal,
degenerate core and the external envelope plays a major role: given these arguments, HBB
cannot be described within a synthetic modeling context, which is one of the main motivations of
the present investigation.

\begin{figure}
\begin{minipage}{0.45\textwidth}
\resizebox{1.\hsize}{!}{\includegraphics{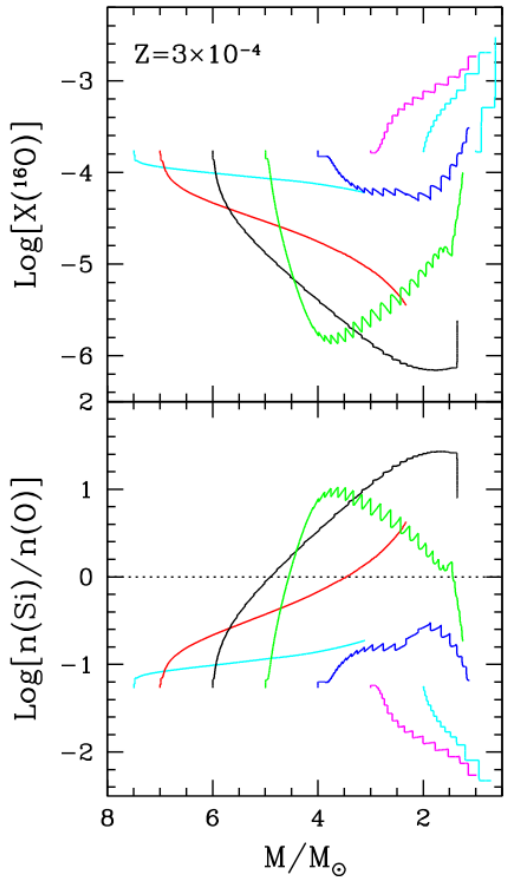}}
\end{minipage}
\vskip-10pt
\caption{{\bf Bottom:} Variation of the surface oxygen of AGB models of various initial
mass during the AGB evolution. Massive models, with mass $M > 5M_{\odot}$, destroy oxygen, 
whereas low--mass stars, with mass below $3M_{\odot}$, owing to repeated third dredge--up 
episodes, increase $^{16}O$. In models with $3M_{\odot} < M < 6M_{\odot}$ oxygen is 
initially destroyed at the surface by HBB, then some production occurs when TDU occurs.
{\bf Top:} evolution of the Si/O ratio in the same models shown in the left panel. The 
dotted line indicates the stage when the number densities of the two species are equal.
}
\label{fhbb}
\end{figure}

HBB can impose a strong variation in the surface chemistry,
because of the proton--capture nucleosynthesis experienced at the bottom of the convective zone,
and the rapidity of convective motions that homogenize the whole envelope: the
chemistry of stars experiencing HBB reflects the equilibria of CNO nucleosynthesis.

The variation of the surface oxygen, shown in the top panel of Fig.~\ref{fhbb},
is a clear indicator of the mechanism dominating the surface chemistry. 
Models with mass $M>5M_{\odot}$ destroy oxygen via HBB; whereas lower mass stars, with 
$M<3M_{\odot}$ increase it via repeated TDU episodes. Note that in models with mass in
the range $(3-5)M_{\odot}$ oxygen is initially destroyed by HBB, then produced by TDU episodes
in the more advanced AGB phases (see last column of Table 2). The oxygen content in the gas 
ejected by super--AGBs is larger than in the matter expelled by massive AGBs: this effect,
discussed in \citet{vd11}, is a consequence of the velocity with which super--AGBs loose their
envelope, the rate at which the mass is lost exceeding the rate at which oxygen
is destroyed. The largest depletion of oxygen, in analogy with higher metallicity models
(see paperI and paperII), is found for the masses at the edge between the AGB and the super--AGB 
regimes, i.e. $M\sim 6M_{\odot}$.

The bottom panel of Fig.~\ref{fhbb} shows the evolution of the surface Si/O ratio; this will 
be important when evaluating the possibility of silicate formation in the atmospheres 
of these objects. In models with $M \geq 5M_{\odot}$ the simultaneous depletion of oxygen 
and production of silicon via the Mg--Al nucleosynthesis \citep{vcd11} makes silicon 
more abundant than oxygen at the surface.

\subsection{The carbon--star stage in low--mass AGBs}
HBB is not active in low--mass stars with $M < 3M_{\odot}$; the only mechanism
able to modify their surface chemistry is TDU, which determines an increase in the
surface carbon. The amount of $^{12}C$ carried to the surface after each thermal pulse depends 
on the extent of the inwards penetration of the convective zone: the more internal is the 
region reached by the base of the envelope, the larger will be the carbon increase. 
Unfortunately, the extent of the third dredge--up is sensitive to many ingredients used in 
calculating the evolutionary sequences, still unknown from first principles \citep{karakas11}. 
The method used to determine the location of the border between convective and radiative 
regions is crucial for the results obtained. While sticking to the Schwarzschild criterium
underestimates the dimensions of the mixed region, the extent of the possible overshoot is still
an object of debate. Other mechanisms, different from convective overshoot,
may further extend the dimensions of the mixed zone: rotational mixing and internal gravity
waves may also favor mixing of layers radiatively stable \citep{stancliffe11}.

Studies focused on the luminosity function of the carbon stars in the Magellanic Clouds
\citep{marigo99, izzard04} calibrated the extent of TDU needed to reconcile the observational 
evidence with theoretical modeling: these investigations provided the minimum core mass
at which TDU begins to be effective in producing carbon stars, 
and the extension of the mixed region. However, we are still 
far from a robust and reliable description, holding for all masses and metallicities.
We decided in this context to avoid any detailed calibration of the
possible extra mixing from convective zones present in the interiors of AGB stars; this
analysis would be harder than for higher metallicity models, from which many more observations
are available. We simply explore how the dust formation process is sensitive to the
depth of TDU. 

The extent of TDU was changed by assuming a variable overshoot from the borders
of the convective shell that forms during the thermal pulse. Accounting for this extra mixing
is known to enhance the strength of the pulse, and to favor the inwards penetration of the convective 
envelope during the TDU \citep{herwig04}. In the following, we compare the results obtained 
with no extra mixing, with those found when values of the overshoot parameter
(see section \ref{refmodel}) $\zeta=0.001, 0.002, 0.005$ are adopted.

\begin{figure*}
\begin{minipage}{0.45\textwidth}
\resizebox{1.\hsize}{!}{\includegraphics{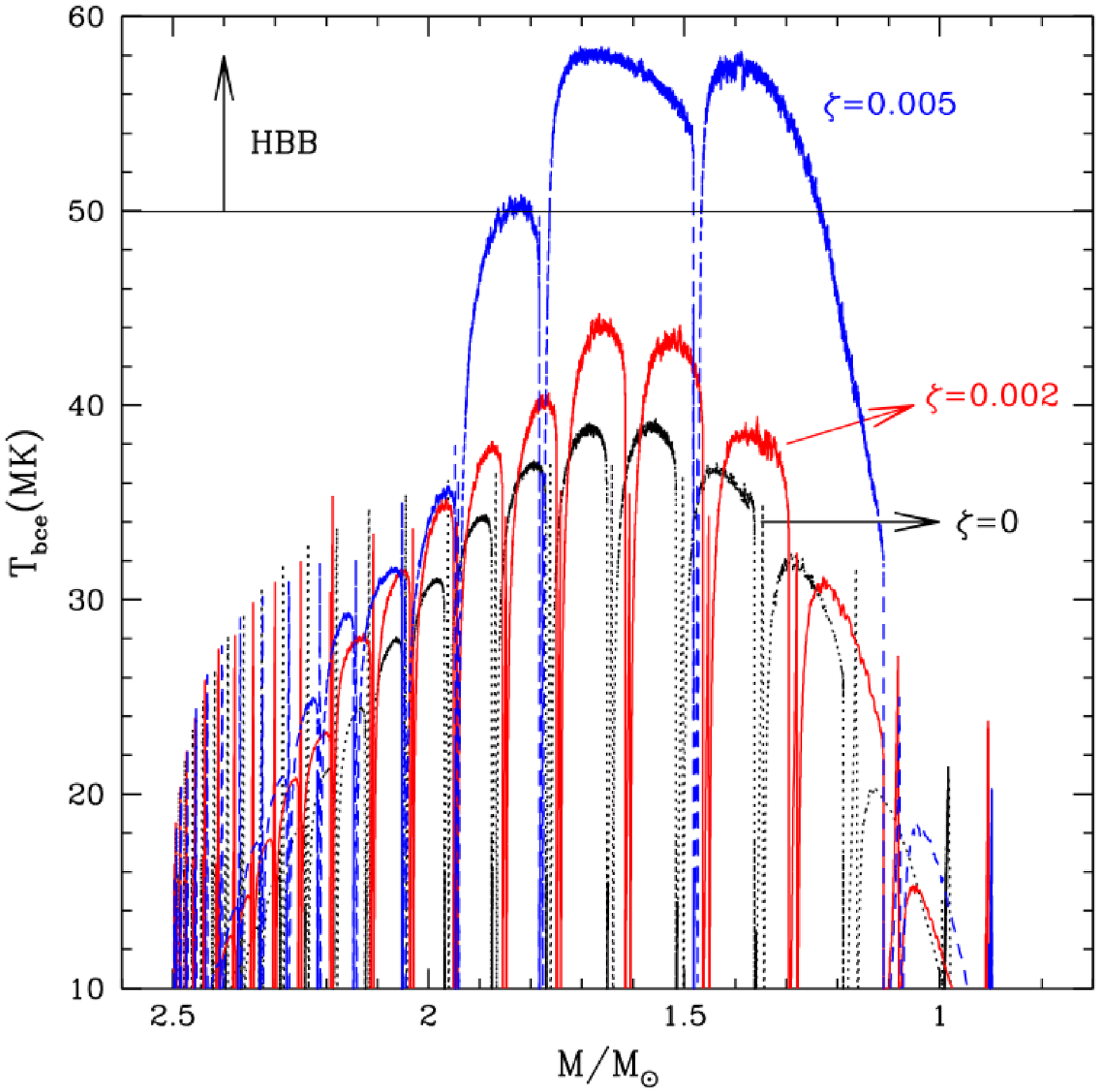}}
\end{minipage}
\begin{minipage}{0.45\textwidth}
\resizebox{1.\hsize}{!}{\includegraphics{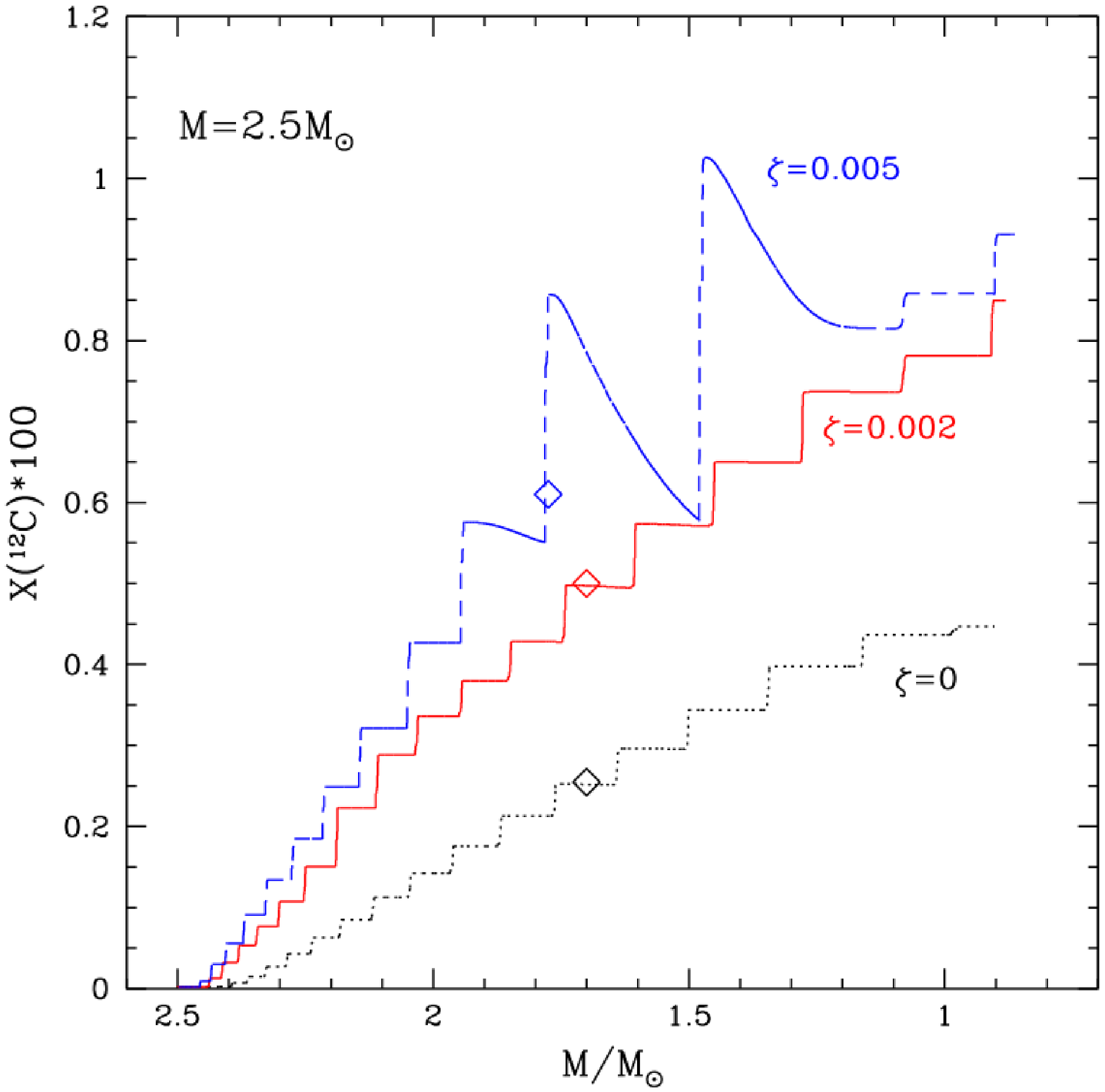}}
\end{minipage}
\vskip-40pt
\caption{{\bf Left}: Evolution of the temperature at the bottom of the convective envelope
of AGB models of $2.5M_{\odot}$ calculated with different prescriptions of the overshoot
from the convective shell: no overshoot (black, dotted), $\zeta=0.002$ (red, solid),
$\zeta=0.005$ (blue, dashed). The thin horizontal line indicates the limit in temperature
leading to HBB conditions, with the depletion of the surface carbon. {\bf Right}: The change 
in the surface carbon in the same models shown in the left panel. The effects of HBB can be 
seen in the decrease in $X(^{12}C)$ during the late inter--pulse phases of the 
$\zeta=0.005$ model. Open diamonds indicate the surface carbon content averaged during the
whole AGB evolution, which corresponds to the quantity shown in Fig.~\ref{cfin}.
}
\label{25msun}
\end{figure*}

The typical evolution of a low--mass star experiencing TDU, in terms
of the progressive increase in the surface carbon, is shown in the right panel of 
Fig.~\ref{25msun}, which refers to a $2.5M_{\odot}$ model. A sudden increase in 
the $^{12}C$ mass fraction follows each TDU episode. The envelope of these models becomes 
more and more enriched in carbon, until the mass left in the envelope is so small that no 
TDU occurs \citep{VW93}.The cases $\zeta=0$ (black, dotted), 
$\zeta=0.002$ (red, solid) and $\zeta=0.005$ (blue, dashed) are shown. 
Fig.~\ref{cfin} shows the average carbon abundance in the envelope during the AGB evolution, 
for models with initial masses $1M_{\odot} \leq M \leq 2.5M_{\odot}$. 

The average carbon in the gas ejected by the $M=2.5M_{\odot}$ 
model is $X(C) \sim 5\times 10^{-3}$, independently of $\zeta$. The reason for this can be 
understood on the basis of the left panel of Fig.~\ref{25msun}, showing the variation of the
temperature at the bottom of the convective envelope. In the $\zeta=0.005$
case (blue, dashed line) the strength of the pulses is enhanced such that eventually
HBB conditions are reached, and the temperature exceeds $\sim 50 \times 10^{6}$K. 
The impact of $\zeta$ 
on the increase in the surface carbon is limited in models with mass approaching the threshold 
for HBB ignition, because the increase in the strength of the thermal pulses eventually 
leads to HBB conditions, which determine a partial destruction of the carbon previously 
accumulated. This can be seen in the right panel of Fig.~\ref{25msun}, showing that in
the $\zeta=0.005$ model the increase in the carbon mass fraction due to TDU is followed
by the destruction of the surface carbon via HBB.

\section{Dust production in low-metallicity AGB stars}
Dust production is strictly correlated with the surface chemical composition of the star.
In our previous investigations we examined the metallicities $Z=0.001$ (paper I) and $Z=0.008$ 
(paper II): in both cases we found that in the winds of stars with an initial mass above the limit 
for the ignition of HBB, i.e. $(2.5-3)M_{\odot}$, conditions are favourable for the formation 
of silicates and very little or no carbon grains are formed. 
Clearly this behaviour cannot be extended to extremely low metallicities, because the 
silicon available in the envelope scales with $Z$; we expect that below a given $Z$ 
silicate formation is scarcely efficient, and the little amount of dust formed is not 
capable of accelerating the wind via radiation pressure on the dust grains.

\begin{table*}
\caption{Dust production by low--mass AGBs with $Z=3\times 10^{-4}$}
\label{tabdust}
\begin{tabular}{ccccc}
\hline
\hline
$M/M_{\odot}$ & $M_C(\zeta=0)/M_{\odot}$ & $M_C(\zeta=0.001)/M_{\odot}$ & $M_C(\zeta=0.002)/M_{\odot}$ & $M_C(\zeta=0.005)/M_{\odot}$  \\
\hline
 & & Z=$3\times 10^{-4}$ & &  \\
\hline
1.0  & -   & -  &  $10^{-4}$    &  $1.5\times 10^{-4}$  \\
1.5  & -   & $10^{-4}$  & $4\times 10^{-4}$ & $4\times 10^{-4}$   \\
2.0  & $1.5\times 10^{-4}$ & $5.8\times 10^{-4}$ & $6\times 10^{-4}$ & $6.4\times 10^{-4}$   \\
2.5  & $3.5\times 10^{-4}$ & $4.5\times 10^{-4}$ & $4.7\times 10^{-4}$ & $5\times 10^{-4}$   \\
\hline
\hline
\end{tabular}
\end{table*}

\subsection{Silicate formation}
To test the possibility that silicates are formed at the metallicity investigated in this
work we focus on $\Gamma$ (Eq.~\ref{eqgamma}): the condition for the wind to be
accelerated by radiation pressure is $\Gamma > 1$, which requires a minimum opacity,
$k_{min}={4\pi cGM\over L}$.

\begin{figure}
\resizebox{1.\hsize}{!}{\includegraphics{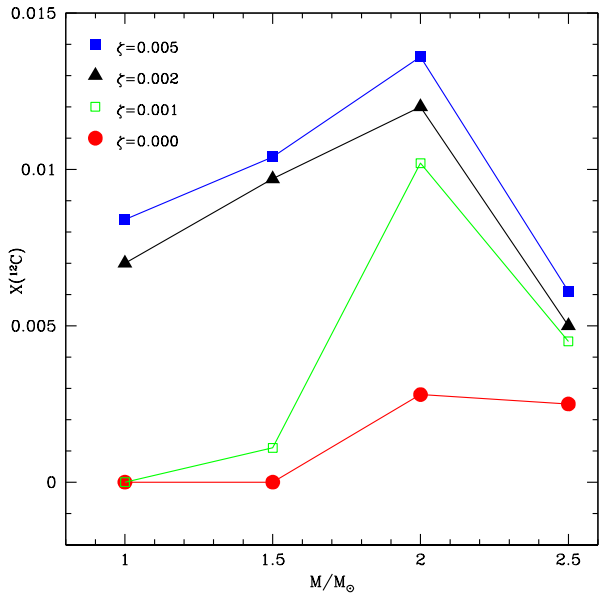}}
\vskip-10pt
\caption{The average abundance of carbon in the gas ejected by low--mass models,
whose surface chemistry is changed by the effects of the Third Dredge--up. The lines
in the figure indicate the results obtained by changing the extent of the overshoot
from the borders of the convective shell formed during the thermal pulse. Stellar
models with $M \leq 1.5M_{\odot}$ and no overshoot never become carbon stars.
}
\label{cfin}
\end{figure}

The most favorable conditions for the formation of silicates are found in the early 
super--AGB phases of models with mass $M\sim 7M_{\odot}$. The high luminosities reached 
by these stars (see Table~\ref{yields} and the
right panel of Fig.~\ref{fagb}) favour large mass loss rates, which, in turn, enhance the
dust formation rate, owing to the increase in the density of the wind (Eq.~\ref{eqmloss}).
Also, as shown in Fig.~\ref{fhbb}, in these models the surface oxygen is never completely
destroyed; this allows the formation of water molecules, present in all the reactions
occurring when the silicates are formed (see Table 1). These conditions are not met during
the evolution of smaller mass stars, whose luminosity is smaller, and that suffer a 
stronger reduction of the surface oxygen (see Fig.~\ref{fhbb}).

We focus on a model of $7M_{\odot}$, in the phase of maximum luminosity ($L=1.3\times 10^5L_{\odot}$),
when it has lost $1M_{\odot}$ (see right panel of Fig.~\ref{fagb}); the surface
temperature is 3700 K. Substituting $M=6M_{\odot}$ and $L=1.3\times 10^5L_{\odot}$ in the
expression given above we find $k_{min}=0.6$cm$^2$/g. On the other hand,  by using the optical 
constants for silicates by \citet{ossenkopf92}, in the unrealistic case that all the silicon 
is condensed into dust, we find $k=0.3$cm$^2$/g, i.e. $k \sim {1\over 2} k_{\rm min}$. 
This result, compared with those from our previous investigations, shows that 
$Z\sim 10^{-3}$ is the threshold metallicity, below which no purely dust--driven silicate 
winds may occur. Some little production of silicates
is still possible in these conditions, but the predictions in this case are less reliable,
because the results depend critically on the assumptions concerning the velocity with which the
wind particles reach the condensation layer.

\subsection{Carbon and SiC dust in low--mass AGBs}
In analogy with the results for higher metallicities, in models with mass below $3M_{\odot}$
dust formation depends on the extent of TDU. The excess of carbon with respect to oxygen
allows formation of solid carbon and silicon carbide.
This condition is reached after a few TPs, provided that TDU is sufficiently deep
to enrich in carbon the surface layers (see right panel of Fig.~\ref{25msun}). 
In the previous phases, as far as the C/O ratio is below unity, only silicate formation is 
possible. The percentage of silicon in the wind is however so low, that the rate of growth of the grains
is very small, the maximum dimensions hardly reaching $~0.05 \mu$m. In agreement with the 
previous discussion, we find that the quantity of dust formed is far below 
the threshold needed to accelerate the wind. This is consistent with the analysis by
\citet{hofner08}, indicating that in the range of masses and luminosities typical of
the low--mass regime investigated here a mimimum dimension of silicates grains of
$\sim 0.3 \mu$m is required to accelerate the wind.

\begin{figure}
%\begin{minipage}{0.45\textwidth}
\resizebox{1.\hsize}{!}{\includegraphics{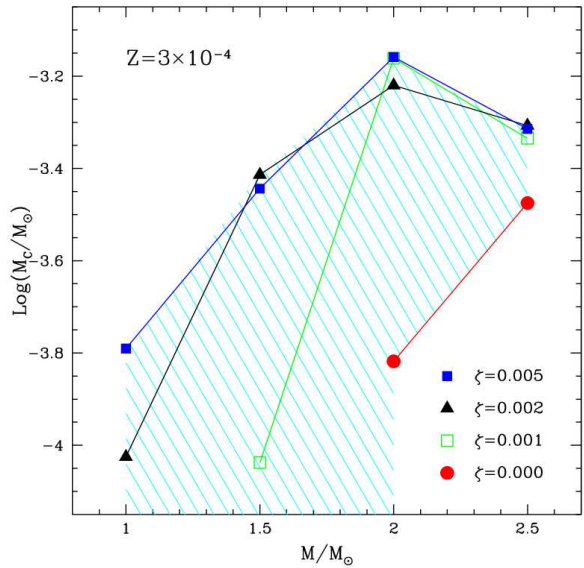}}
%\end{minipage}
\vskip-50pt
\caption{Mass of carbon dust formed in the winds of AGB stars of various mass. The different
symbols refer to models calculated with different extent of the extra mixing from the
borders of the convective shell which forms during the thermal pulse. The shaded area
indicates the degree of uncertainty associated with the carbon dust.
}
\label{fmasscar}
\end{figure}

Table~\ref{tabdust} reports the mass of solid carbon produced 
by AGB models of different initial mass, calculated with $\zeta=0, 0.001, 0.002, 0.005$.
The same data are shown in Fig.~\ref{fmasscar}.

Models with mass below $2M_{\odot}$ become carbon stars only for $\zeta \geq 0.002$. For a 
given $\zeta$, the mass of carbon dust produced increases with the mass of the star, because
more massive models experience more mass loss and more thermal pulses, thus reaching 
higher carbon abundances. The trend in Fig.~\ref{fmasscar} reflects the behavior of the carbon 
content in the gas ejected by AGBs, shown in Fig.~\ref{cfin}. 
The $2.5M_{\odot}$ model is found to produce a smaller amount of
carbon, which can be explained based on the discussion in previous sections, concerning the
HBB ignition in this model.

Fig.~\ref{fmasscar} shows that the mass of carbon dust produced becomes approximately 
independent of $\zeta$, for $\zeta > 0.001$. 
The main reason for this behaviour is the rapid acceleration
experienced by the wind when carbon dust is produced: the larger is the content of carbon dust,
the larger is the acceleration of the wind, which prevents additional formation of dust. 
In addition, models more enriched in carbon are those experiencing the deepest TDU, thus evolving 
at smaller core masses: their luminosity, and consequently their mass loss rate, is 
smaller, such that the density of the gas in the wind is also reduced.

\begin{figure*}
\begin{minipage}{0.32\textwidth}
\resizebox{1.\hsize}{!}{\includegraphics{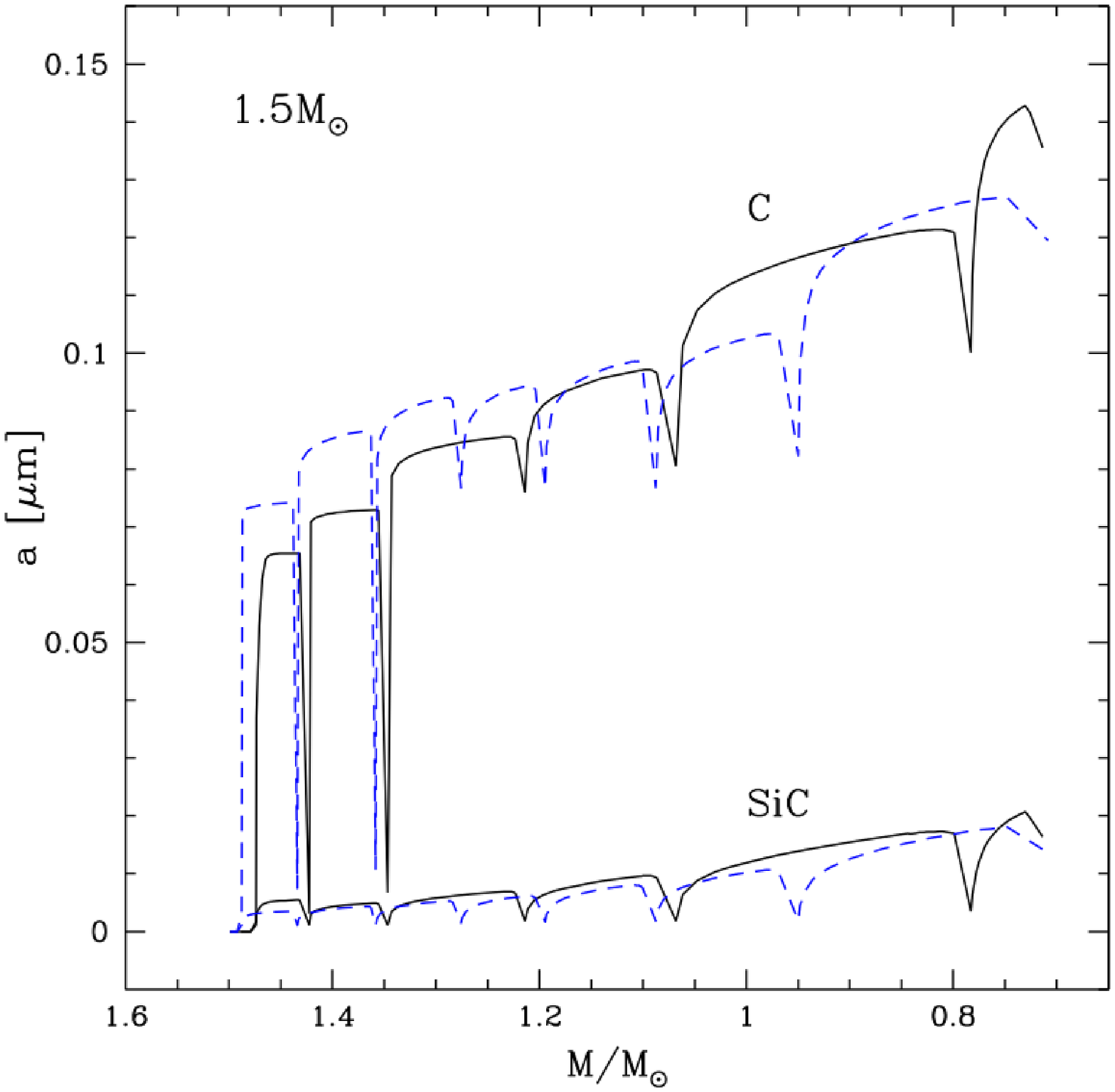}}
\end{minipage}
\begin{minipage}{0.32\textwidth}
\resizebox{1.\hsize}{!}{\includegraphics{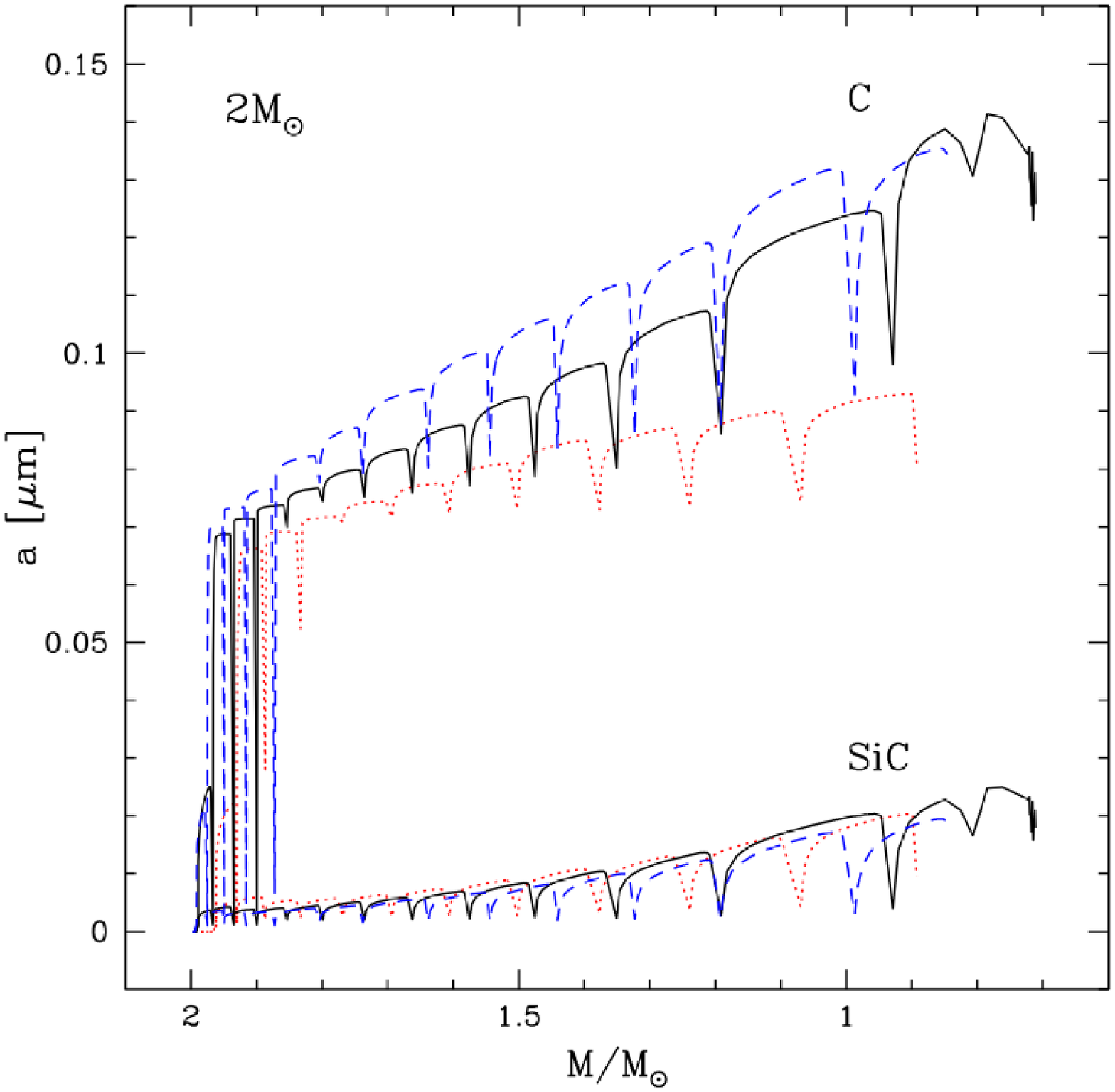}}
\end{minipage}
\begin{minipage}{0.32\textwidth}
\resizebox{1.\hsize}{!}{\includegraphics{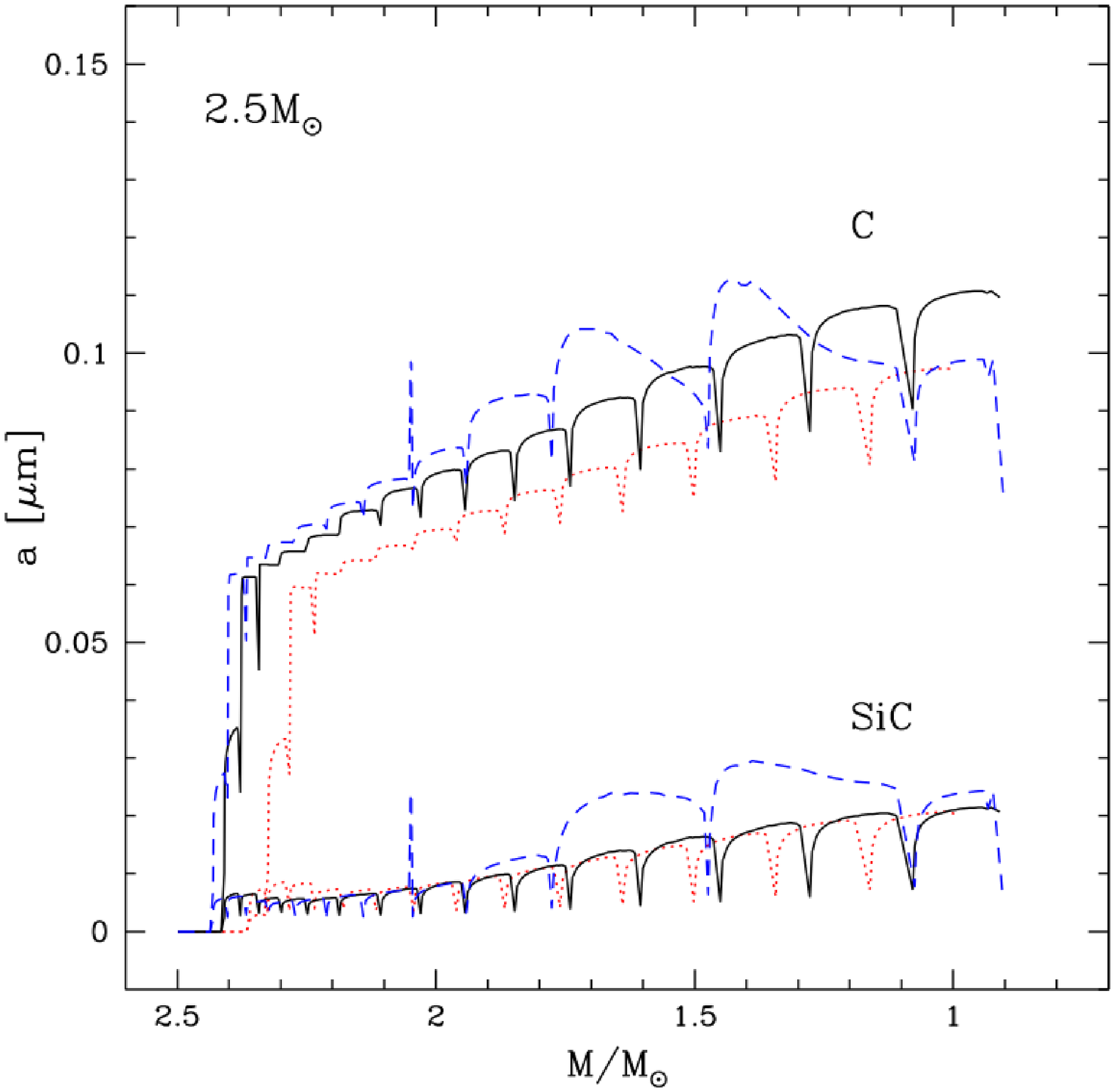}}
\end{minipage}
\vskip-40pt
\caption{The dimension of carbon and SiC dust grains formed around models with initial 
mass $1.5M_{\odot}$ ({\bf Left}), $2M_{\odot}$ ({\bf Center}), $2.5M_{\odot}$ ({\bf Right}) 
during the AGB evolution. Different lines refer to the assumptions concerning the 
extra mixing from the borders of the convective shell which forms during the thermal 
pulses, indicated by the parameter $\zeta$ (see text for details). Dotted: no--overshoot; 
solid: $\zeta=0.002$; dashed: $\zeta=0.005$.}
\label{facar}
\end{figure*}

Fig.~\ref{facar} shows the variation of the size of C and SiC grains during the AGB evolution 
of stars of initial mass $1.5M_{\odot}$, $2M_{\odot}$, $2.5M_{\odot}$; the 
different lines indicate the $\zeta$ adopted in the computations.

The size of the SiC grains is independent of $\zeta$, because the key element, Si, is
unaffected by TDU. Owing to the scarcity of silicon available, the production 
of SiC is modest, the size of SiC grains being $a_{\rm SiC} < 0.02 \mu$m. The amount of
SiC formed increases with mass, because more massive models experience a larger mass loss
rate, that favours gas condensation.

Iron is scarcely produced in all models, as a consequence of the small iron content of
the winds of AGBs at these low metallicities. This result is consistent with previous
investigations by \citet{fg06} and \citet{paperI}, indicating that the iron dust produced
by AGBs diminishes with decreasing metallicity. This picture could change at higher
metallicities, as suggested by the analysis by \citet{mcdonald11}, focused on dust
production in stars belonging to the Globular Cluster 47Tuc. The authors find that iron
is the main component of the dust content surrounding the two brightest sources,
suggesting a favourite channel for iron production compared to silicates. Detailed models
of low--mass stars at higher metallicities are required to be compared with these findings.

The growth of carbon grains varies with $\zeta$, because the surface 
carbon enrichment changes according to the depth of TDU, as shown in Fig.~\ref{cfin}. 
The grain size increases during the evolution, not only because more and more carbon is
brought to the surface, but also because the luminosity, hence the mass loss rate,
increases as the core grows in mass. While a minimum value of $\zeta$ is required 
to allow TDU to operate efficiently, the results in terms of the size of carbon grains
formed become insensitive to $\zeta$ as far as a sufficient amount of carbon, 
$X(C) > 5\times 10^{-3}$, is present in the envelope. 

A calibration of $\zeta$ is beyond the scopes of the present work. However, in the analysis
of the carbon grain size formed, we may disregard the no--overshoot assumption, given the 
physical inconsistency, and the severe problems met when comparing the results with the 
observations of stars in the Magellanic Clouds \citep{marigo99, izzard04}. Limiting our 
attention on the $\zeta \neq 0$ 
cases, we conclude that the surroundings of low--mass, low--metallicity AGBs are expected 
to host carbon grains, whose dimensions are in the range $0.08 \mu$m $< a_C < 0.12 \mu$m
(see Fig.~\ref{facar}). These results are approximately independent of the initial mass
of the star.

\begin{figure}
\begin{minipage}{0.45\textwidth}
\resizebox{1.\hsize}{!}{\includegraphics{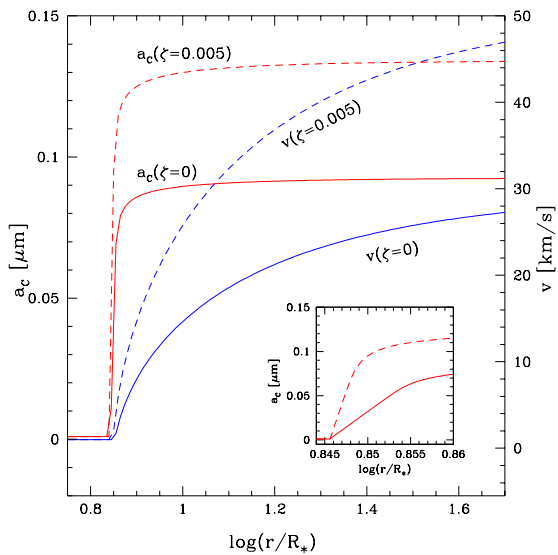}}
\end{minipage}
\vskip-40pt
\caption{Radial profiles of velocity (blue) and size of carbon grains (red) in
the wind expanding from the surface of a $2M_{\odot}$ model calculated with no extra mixing 
(solid) and with $\zeta=0.005$ (dashed). The internal panel, focused on the dust forming 
region, shows only the grains size.
}
\label{fvela}
\end{figure}

The reason for this is the rapid acceleration of the wind favoured by the condensation
process; the increase in the velocity favours a rapid decrease in the density of the gas, 
that inhibits further formation of dust. In Fig.~\ref{fvela} we compare the structure of
the wind of two models of $2M_{\odot}$, calculated with $\zeta=0$ and $\zeta=0.005$. The
figure refers to an evolutionary phase when the stars have lost $1M_{\odot}$, and the
surface carbon abundances are $X_C=0.003$ ($\zeta=0$) and $X_C=0.014$ ($\zeta=0.005$).
The radial profiles of the grain size (red) and velocity (blue) are shown.
Solid tracks refer to the no--overshoot models, whereas dashed lines indicate the $\zeta=0.005$
case.

Dust condensation begins at a distance of $\sim 10R_*$ from the stellar center, and continues
until the velocity of the wind reaches $\sim 10$km/sec. The acceleration of the wind is smaller
in the $\zeta=0$ case, which allows dust formation to occur in the expanding wind up to more
external regions. As clearly evident in the internal panel of Fig.~\ref{fvela}, showing in
more details the dust--forming region, this effects partly compensates the difference in the
carbon available in the winds, reducing the gap between the size of the carbon grains 
formed. 

The optical properties of solid carbon, its ability to absorb radiation and accelerate
the wind, make the description of the solid carbon growth only mildly dependent on the 
quantity of carbon in the stellar envelope, thus rendering more robust the results obtained,
in terms of the dimension of the grains formed.

The possibility of comparing these results with the observations is undermined by the
scarcity of data of AGB stars in metal poor environments. A promising field in this context
is the increasing amount of available spectra of AGBs in globular clusters, from which the species
and the quantity of dust formed can be deduced. Because masses evolving in Globular Clusters
are surely below 1M$_{\odot}$, this approach offers the only opportunity to test our 
description at the lower end of the range of masses investigated, whereas no information 
can be obtained for models of higher mass. An example of these studies is the collection
of Spitzer data presented by \citet{sloan10}, showing a positive trend with $Z$ of the 
amount of dust formed in the circumstellar shells of AGB stars. 

While stars in the sample presented in \citet{sloan10} are more metal rich than the models
presented here, the two sources belonging to NGC 4372, analyzed in \citet{mcdonald13},
have a similar metallicity ([Fe/H]$=-2.2$), which offers the opportunity of a more detailed 
comparison. From the spectra, the authors argue that the two stars are oxygen--rich, and 
dusty. The luminosity of the two stars is slightly below the value found in the $1$M$_{\odot}$ 
model presented here (see Table 2), in agreement with the smaller mass ($\sim 0.9M_{\odot}$) 
expected to evolve in the AGB of this cluster. Fig.\ref{fmasscar} shows that some 
carbon--dust production is expected in these low--mass 
objects, provided that some extra mixing from the convective zones is adopted. This is 
however not in contrast with the fact that the two stars examined by \citet{mcdonald13} 
are oxygen--rich, because even in case that a large extra mixing is used, the model
achieve the C--star stage only in the lates evolutionary phases, while the surface
remains oxygen--rich for most of the duration of the AGB phase.

\subsection{Dust mass produced by AGB stars: the trend towards metal--poor chemistries}
As discussed above, the quantity of silicon present in low metallicity AGBs is so small to prevent 
the formation of silicates, if not in extremely small quantities. Even in the super--AGB regime,
where the strong mass loss might favor the condensation process, the density of silicon
in the wind is too small to allow dust production. Our AGB models experience stronger HBB
than other literature results \citep{karakas10}, and are thus those most favorable to the formation of 
silicates. This makes these results more general, allowing us to conclude that 
$Z \approx 10^{-3}$ is the threshold limit in metallicity, below which no meaningful 
silicate formation occurs in the winds of AGBs.

The shortage of silicon in the wind also affects the formation of SiC, whose production
rate scales with the silicon density.

\begin{figure}
\begin{minipage}{0.45\textwidth}
\resizebox{1.\hsize}{!}{\includegraphics{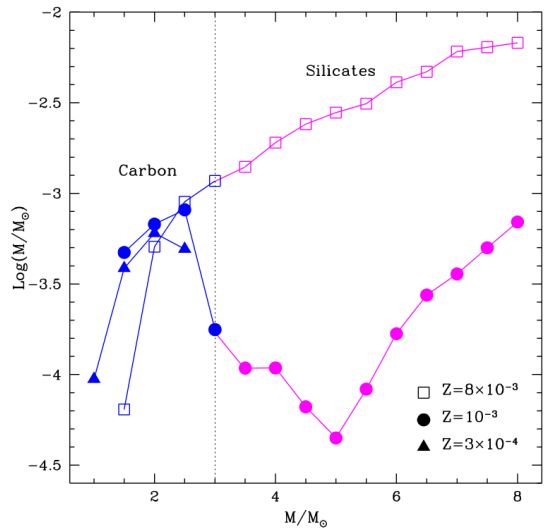}}
\end{minipage}
\vskip-40pt
\caption{The mass of dust produced by AGB models of different initial mass,
at various metallicities. Full points indicate results for $Z=10^{-3}$ 
published in paper I, open squares refer to the $Z=8\times 10^{-3}$
metallicity presented in paper II, whereas full triangles refer to the present
investigation. The models in the low--mass regime were calculated by assuming 
an extra mixing $\zeta=0.002$ from the convective shell which forms during each TP.
The thin, dotted line separates models producing silicates from those producing
carbon dust.
}
\label{fmassmet}
\end{figure}

The comparison of the results obtained in the present investigations with those presented 
in paper I and paper II allows us to infer how the mass and composition of dust produced by 
AGBs change with metallicity. Fig.~\ref{fmassmet} shows the mass of dust produced by stars 
with masses in the range $1 M_{\odot} \le M \le 8 M_{\odot}$ and metallicities $Z=10^{-3}$ 
(paper I), $Z=8\times 10^{-3}$ (paper II), and $3\times10^{-4}$ (present study).

Stars with mass above $\sim 3M_{\odot}$ produce silicates, in quantities scaling with $Z$. 
Low--mass stars produce mainly carbon dust, although some silicate formation occurs,
in quantities increasing with metallicity, owing to the larger silicon available. As far
as the C--star stage is reached, the carbon--dust formed exceeds by far the previous
production of silicates, because of the large excess of carbon with respect to oxygen
favoured by repeated TDU episodes. Most of the mass of dust shown in Fig.~\ref{fmassmet},
produced in the low--mass regime, is under the form of carbon--dust.

The effect of metallicity on the carbon formed 
is not trivial, because the carbon dredged to the surface during the TDU is produced in 
the He--burning shell, and is thus independent of $Z$. In low--$Z$ models TDU is more efficient 
\citep{boothroyd88}, thus their surface carbon is larger. This effect is compensated by the 
cooler surface temperatures at which higher--$Z$ models evolve: the dust--formation zone is closer 
to the stellar surface, thus the condensation process is more efficient. Moving to smaller 
metallicities the range of masses involved in the solid carbon formation becomes 
progressively smaller, because the larger core masses favour HBB conditions for smaller 
and smaller initial masses. For masses $M < 2M_{\odot}$, lower $Z$ models produce more
dust, because they reach earlier in their evolution the carbon star stage.

{The results shown in Fig.~\ref{fmassmet} indicate that while the production of silicates 
turns out to be extremely dependent on metallicity, the carbon--dust produced show only
(if any) a modest dependence on $Z$, but the lower and upper limits of the masses involved.
These findings are in agreement with the results presented by \citet{sloan12}, showing that
that the amount of amorphous carbon produced is fairly independent of metallicity, for
carbon stars observed not only in the Galaxy and in the Magellanic Clouds, but also in 
other galaxies of the Local Group.}

We plan to further extend these results to smaller metallicities, but the general pattern
can be inferred from the results found here and from the properties of AGB stars at extremely
small $Z$. In the high--mass domain, silicate formation becomes harder with decreasing $Z$, 
because of the smaller amount of silicon available and the stronger HBB, which triggers the
destruction of surface oxygen. For masses $M \geq 3M_{\odot}$ HBB is 
expected to occur down to $Z=10^{-8}$ \citep{lau09}, which prevents 
the formation of carbon--type dust. 

At $Z$ below $10^{-4}$, low--mass AGBs evolve differently 
\citep{lau09, campbell08}. The ignition of He--burning, both in the core or in a shell, is 
accompanied by the development of an extended convective zone, which, owing to the small 
height of the entropy barrier, penetrates the H--rich region. The following inwards 
penetration of the external envelope allows mixing of processed material in the interior 
with the surface, which is thus enriched in CNO elements. The composition of the winds of 
these stars depends on which mechanism is more efficient in changing the surface chemistry.
Stars with mass above $\sim 1M_{\odot}$ experience both HBB and TDU; although dual shell
flash is also likely to occur, the change in the surface chemistry is mostly determined
by the combined effects of TDU and HBB: the former favors an enrichment in the surface carbon 
after each TP, the latter destroys part of this carbon and produces nitrogen. 
As a result, the average carbon mass fraction barely reaches $X(C)=5\times 10^{-4}$, 
thus preventing the formation of meaningful quantities of carbon grains. 
Therefore, we can expect that when $Z \le 10^{-4}$, dust production is limited to low--mass 
stars, whose evolution is not influenced by HBB and whose surface carbon may survive to
proton fusion.

\subsection{The uncertainties in the modelling}
Before addressing the implications of the results presented, we discuss the
robustness of these findings. The treatment of convection and mass loss are known
to have a significant impact on AGB modelling, both on the chemical and physical sides
\citep{vd05a, vd05b, vm10}. However, at odds with previous explorations based on higher
metallicities \citep{paperI, paperII}, dust production in the high--mass domain is not
sensitive to these uncertainties at these low Z's: the amount of silicon 
in the wind is so scarce, that no meaningful silicate--type dust is expected, far below 
the quantities needed to accelerate the wind. 

The choice of the convection modelling influences the upper limit of the masses
producing carbon--type dust: if a less efficient description of convection was adopted,
the threshold limit (in mass) to experience HBB would shift $\sim 0.5M_{\odot}$ upwards,
which would extend the range of masses involved in carbon production up to
$\sim 3M_{\odot}$. 

The description of mass loss could potentially influence the physical evolution of
low--mass stars: the mass loss rate determines the number of thermal pulses experienced,
thus the extent of the surface carbon enrichment. Yet, we do not expect a strong
sensitivity of the amount of carbon--dust formed to the mass loss description for the
following reasons: a) a higher carbon enrichment favours a rapid cooling and expansion 
of the outer layers of the stars, with the consequent increase in the mass loss rate
experienced \citep{marigo02}, which prevents many more TDU episodes to occur; b) as
discussed previously, the size of the carbon grains formed are not strongly sensitive 
to the amount of carbon present in the wind, owing to the high opacity of carbon grains,
that favour a strong acceleration of the winds as soon as they begin to form.

Concerning the wind modeling, the robustness of the present results must be confronted 
with a more complete description, which accounts for the presence of shock waves. Dust 
formation would be favoured by the shock, because compression of the gas lifted upwards
would increase the densities of the key species, and dense parcels of gas would be
pushed to cool regions distant from the central star, where dust is expected to be formed 
in great quantities.

In oxygen--rich environments, a further effect due to the presence of shock waves might 
be the deviation from chemical equilibrium in the gas phase, with the consequence that
carbon is not entirely locked into CO molecules, part of it remaining available for
carbon--dust formation. This scenario could lead to the simultaneous formation of
carbon and silicates grains in the circumstellar shells of oxygen--rich AGBs, and would 
provide an alternative mechanism for a dust driven mass loss for M stars
\citep{hofner07}.

\section{Conclusions and implications for cosmic dust enrichment}

We calculate the dust formed around AGB and super--AGB stars of metallicity $Z=3\times 10^{-4}$,
the lowest metallicity for which dust production by low and intermediate mass stars has been
considered so far.
This study complements previous investigations that were limited to metallicities $Z=10^{-3}$ 
and $Z=8\times 10^{-3}$. Stellar evolution is followed by means of a full integration of the 
equations of stellar structure, allowing to describe self--consistently the Hot Bottom 
Burning phenomenon.

We find that for stars with initial metallicity $Z = 3\times 10^{-4}$:

\begin{itemize}
\item In the high--mass domain, with $M \geq 3M_{\odot}$, strong HBB prevents the formation of
carbon stars. The scarcity of silicon in the wind and the strong
destruction of the surface oxygen in the most massive models lead to a negligible 
formation of silicates, not sufficient to accelerate the wind via radiation pressure
on dust grains.

\item Dust production is limited to the formation of solid carbon grains around low--mass stars, 
with initial mass $M \leq 2.5M_{\odot}$. The results depend on the way TDU is modelled, 
but as far as a minimum abundance of carbon of the order of $X_C \sim 0.005$ is reached in the surface 
layers, the results are approximately independent of the extent of TDU. This is a consequence of  the balance
between larger carbon abundances reached by models experiencing the deepest
dredge--up and their lower core masses, that lead to smaller mass loss rate. 

\item Under these conditions, the carbon dust produced is $2\times 10^{-4}M_{\odot} < M < 6\times 10^{-4}M_{\odot}$ with grains
radii in the range $0.08 \mu$m $< a_C < 0.12 \mu$ m. 

\item Compared to higher metallicity models, low--mass stars with $M \sim 1-1.5M_{\odot}$
produce more dust because they reach more easily the carbon--star stage. Conversely,
for $M > 1.5M_{\odot}$ higher $Z$ models produce dust more efficiently, owing to their cooler surface
layers.

\item These results can be extrapolated to even more  metal--poor AGBs. 
No dust is expected in the high--mass domain, dominated by HBB, because of the scarcity of 
silicon available. At low Z's HBB is present even in low--mass stars, below $2M_{\odot}$. 
Although it is not sufficiently strong to burn oxygen, still it destroys part of the
carbon accumulated by TDU. The scarcity of carbon in the envelope prevents 
the formation of carbon grains, if not in small quantities.
\end{itemize}

These results have interesting implications for the contribution of AGB stars to dust enrichment.
The grid of stellar models that we have presented in paper I, II and in the present study, suggest
that for $Z = 8 \times 10^{-3} = 0.4 Z_{\odot}$, the metallicity of the Large Magellanic Cloud 
(LMC), AGB stars can contribute to dust enrichment of the ISM on relatively short time-scales,
$\approx   40$~Myr, comparable to the evolutionary time of a $8 M_{\odot}$ star. According
to our results, the enrichment is initially limited to silicate dust, and carbon dust is released
on longer timescales, $t > 300$~Myr, comparable to the evolutionary time of a  $3 M_{\odot}$ star.
For initial metallicities in the range $3 \times 10^{-4} < Z \le 10^{-3}$ or
$0.015 Z_{\odot} < Z \le 0.05 Z_{\odot}$, the production of silicates by the most massive AGB stars
is strongly suppressed and low-mass stars can mostly contribute to carbon dust enrichment 
when $t > 300$~Myr.

It is important to stress that these results and the dependence of the predicted
AGB and super--AGB dust yields on the initial stellar metallicity reflect the complex evolution
of these stars and can not be appreciated when synthetic stellar models are adopted.
Indeed, our results partly contradict previous claims that AGB stars can always 
contribute to carbon dust enrichment, independently of their initial mass and metallicity,
at least in the metallicity range, $5\times 10^{-2} Z_{\odot} \le Z \le 1 Z_{\odot}$
(Ferrarotti \& Gail 2006; Zhukovska et al. 2008). 

More importantly, our study suggests that the dust yields computed for stars with initial 
metallicity of $Z \sim 10^{-2} Z_{\odot}$ can not be extrapolated to lower metallicity as 
the contribution of AGB stars to dust enrichment can be safely neglected when the 
metallicity of the stars is $Z < 10^{-4} (5\times 10^{-3} Z_{\odot}$). Hence, our results 
imply that at these low metallicities, supernovae are left as the only viable stellar dust 
sources (Todini \& Ferrara 2001; Nozawa et al. 2003, 2007; Bianchi \& Schneider 2007;
Cherchneff \& Dwek 2010).

\section*{Acknowledgments}
The authors are indebted to Paola Marigo, for the computation of the low--temperature
opacities in the C--star regime, by means of the AESOPUS tool. RS acknowledges that 
the research leading to these results has received funding from the European Research Council 
under the European Union's Seventh Framework Programme (FP/2007-2013) / ERC Grant 
Agreement n. 306476. It is a pleasure to thank the referee, Ian McDonald, for the careful
reading of the manuscript, and for the many comments and suggestions, that helped
improving the quality of this work.

\end{document}